\documentclass[aps,prl,twocolumn,showpacs]{revtex4}
\usepackage{graphicx}
\usepackage{amsmath}
\usepackage{amssymb}
\usepackage{amsfonts}
\usepackage{bm}

\newcommand{\bea}{\begin{eqnarray}}
\newcommand{\eea}{\end{eqnarray}}
\newcommand{\be}{\begin{equation}}
\newcommand{\ee}{\end{equation}}
\newcommand{\rr}{\mathbf{r}}
\newcommand{\kk}{\mathbf{k}}

\begin{document}

\title{Ground state energy of the two-dimensional weakly interacting Bose gas: 
First correction beyond Bogoliubov theory}

\author{Christophe Mora$^{1}$ and Yvan Castin$^{2}$}
\affiliation{Laboratoire Pierre Aigrain$^{1}$
and Laboratoire Kastler Brossel$^{2}$, \'Ecole Normale
Sup\'erieure and CNRS, Universit\'e Denis Diderot 7$^{1}$ and UPMC$^{1,2}$, 24 rue Lhomond, 75005 Paris, France}

\begin{abstract}
We consider the grand potential $\Omega$ of a two-dimensional
weakly interacting homogeneous Bose gas at zero temperature. 
Building on a number-conserving
Bogoliubov method for a lattice model in the grand canonical ensemble,
we calculate the next order term as compared to the Bogoliubov
prediction, in a systematic
expansion of $\Omega$ in powers of the parameter measuring the
weakness of the interaction. Our prediction is in very good agreement
with recent Monte Carlo calculations.
\end{abstract}

\date{\today}

\maketitle

Recent experimental progress with ultracold atoms
has renewed the interest in the two-dimensional weakly interacting
Bose gas \cite{Jean_BKT,Phillips_BKT}.
In view of a comparison to future experimental results
on the equation of state of the gas at low temperatures, 
this raises the question of the
accuracy of existing theoretical work \cite{revue1,revue2}.
Since the pioneering works of Schick \cite{Schick} 
and Popov \cite{Popov}
on the energy of the weakly interacting Bose gas 
in two dimensions, several recent predictions have
been obtained. In mathematical physics, it was proved that 
Schick's formula for the ground state energy
is asymptotically exact in the limit of vanishing
density \cite{Lieb}. 
Numerically, very precise Monte Carlo calculations of the ground
state energy have been performed \cite{Giorgini,Astra}.
Analytically, Popov's result was confirmed
by a Bogoliubov type theory \cite{Mora} (and by Monte Carlo 
calculations~\cite{Astra})
but several attempts to calculate analytically
the energy beyond Popov's result have led to non-identical predictions
\cite{Cherny,Andersen,Pricoupenko}.

The most systematic among the theoretical approaches
are those relying on an expansion of the energy in powers 
of a small parameter.
This is the case of the approaches \cite{Popov,Mora},
which have led to the equation of state \cite{info_Popov}
\be
\rho \simeq \frac{m\mu}{4\pi\hbar^2} 
\ln\left(\frac{4\hbar^2}{m\mu a^2 e^{2\gamma+1}}\right)
\label{eq:popov-mora}
\ee
where $m$ is the mass of a boson, $a>0$ is the two-dimensional
scattering length among the particles, $\mu$ is the chemical potential
and $\gamma=0.57721566\ldots$ is Euler's constant. 
Remarkably Eq.(\ref{eq:popov-mora}) is {\sl universal}, i.e.\ it depends on the interaction
potential through the scattering length only.
One obtains from (\ref{eq:popov-mora}) 
the grand potential $\Omega=E-\mu N$ in the thermodynamic limit, 
where $E$ is the gas energy 
and $N$ the atom number, by a simple integration over $\mu$ since
$N=-\partial_\mu \Omega$:
\be
L^{-2} \Omega(\mu) \simeq
-\frac{m\mu^2}{8\pi\hbar^2}\left[
\ln\left(\frac{4\hbar^2}{m\mu a^2 e^{2\gamma+1}}\right) + \frac{1}{2}
\right],
\label{eq:omega2}
\ee
where $L^2$ is the surface of the gas.
The small parameter
\be
\epsilon(\mu) = \frac{1}{\ln[4 \hbar^2/(\mu m a^2 \, e^{2\gamma +1})]}
\ee
is apparent in \eqref{eq:omega2}.

\begin{figure}[htb]
\includegraphics[width=8cm,clip]{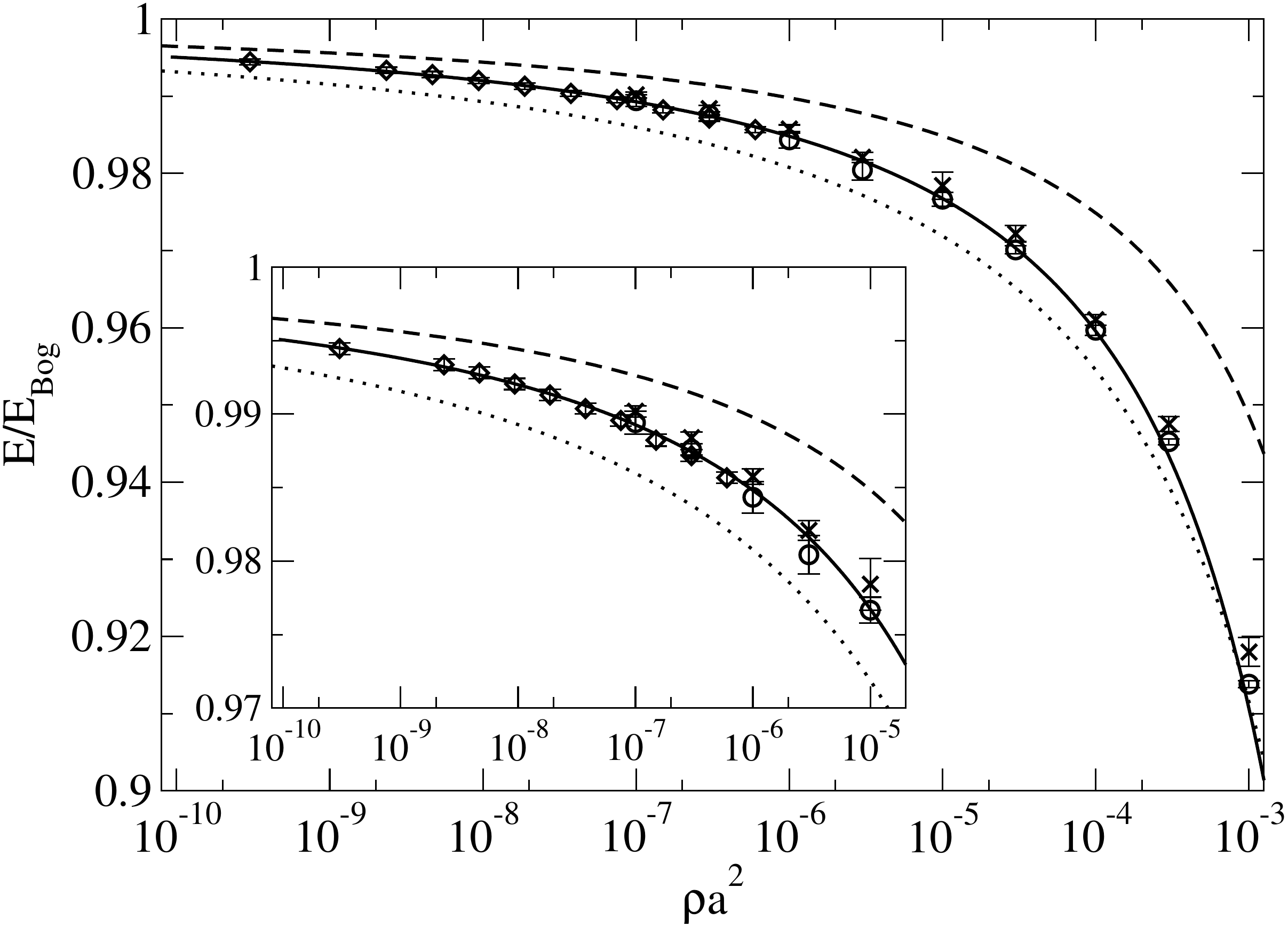}
\caption{Ground state energy $E$ of a two dimensional Bose gas,
as a function of the gas density, and in units
of the Bogoliubov prediction $E_{\rm Bog}$ resulting from
(\ref{eq:omega2}).
Solid line: energy obtained from the beyond Bogoliubov analytical prediction 
(\ref{eq:final}) derived in this work.
Dashed line: analytical prediction of \cite{Cherny} taking the exact expression
$f(u)=u+u^2/2+2u^2 e^{2/u}\mathrm{Ei}\,(-2/u)$ 
of the function $f$ introduced in \cite{Cherny}, where $\mathrm{Ei}$ 
is the function
exponential integral.
Dotted line: analytical prediction of \cite{Pricoupenko}
in the form \cite{param}.
Plotting symbols with error bars: 
numerical results of \cite{Giorgini}, for interactions
given by hard disks (crosses) 
and by soft disks (circles); numerical results of
\cite{Astra}, for dipolar interactions (diamonds). We have restricted
the results of \cite{Giorgini}
to their universal range, that is to low enough values of $\rho a^2$
such that the hard disks and soft disks models give 
the same values of the energy within the error bars.
For the same reason, we have restricted the results of \cite{Astra}
to $\rho a^2< 10^{-6}.$
The inset is a magnification.
\label{fig:comp}
}
\end{figure}

In the vanishing density (or chemical potential) limit, 
the prediction (\ref{eq:popov-mora}) can be checked to be asymptotically
equivalent to Schick's formula, as it should be. 
In a further expansion of the energy
in terms of the density, the precise value
of the constant under the logarithm in (\ref{eq:popov-mora}) eventually
matters. In particular, a careful account of the low-energy two body
$T$-matrix is essential to derive this constant~\cite{info_Popov}.
Its value turns out to agree with recent Monte Carlo results \cite{Astra},
it however differs from the prediction of Ref.~\cite{Andersen}.
For not extremely small values of the density, a significant
deviation is observed between the energy deduced from Eq.~(\ref{eq:omega2})
and Monte Carlo results~\cite{Giorgini,Astra}, see in Fig.\ref{fig:comp} 
the fact that the symbols significantly deviate from unity.
Furthermore, this deviation is not accounted for by the beyond-Bogoliubov
theories of Refs.~\cite{Cherny,Pricoupenko}, see in Fig.\ref{fig:comp}
the fact that the symbols significantly deviate from the dashed
and dotted lines.

In the present work we extend the Bogoliubov method~\cite{Mora} as in \cite{Wu}
and we go
 one step further than 
Eq.~(\ref{eq:omega2}) in the expansion of the grand potential in powers of
$\epsilon(\mu)$. We obtain in the 
thermodynamic limit:
\be
L^{-2} \Omega(\mu) = -\frac{m\mu^2}{8\pi\hbar^2}\left[
\frac{1}{\epsilon(\mu)} + \frac{1}{2}  +
\frac{8I}{\pi} \epsilon(\mu) + \ldots\right],
\label{eq:final}
\ee
where the numerical constant $I$ is given by a multiple integral
that we have evaluated numerically:
\be
I\simeq 1.0005\ldots
\label{eq:I}
\ee
Since the extra term that we have found with respect to (\ref{eq:omega2})
is indeed $o(1)$, this analytically
confirms that the numerical constant inside the logarithm
in (\ref{eq:popov-mora}) is the correct one. Furthermore, the inclusion
of the extra term leads to a now satisfactory agreement with
the numerical results of \cite{Giorgini,Astra}, see in Fig.\ref{fig:comp}
the agreement of the solid line with the plotting symbols.

\noindent{\sl Our model:}
In a first stage we still consider a general value $d$ of the space dimension.
As a regularization scheme to treat ultraviolet divergences,
 we use a lattice model~\cite{Mora} to represent the 
interacting Bose gas, with the grand canonical Hamiltonian
\be
H = \sum_{\rr} \ell^d \hat{\psi}^\dagger \left[ -\frac{\hbar^2}{2m}\Delta
-\mu \right] \hat{\psi}
+ \frac{g_0}{2} \sum_{\rr} \ell^d \hat{\psi}^\dagger \hat{\psi}^\dagger
\hat{\psi}\hat{\psi},
\ee
where the limit $\ell \to 0$ is eventually taken to recover the continuous gas.
The discrete positions $\rr$ run over the lattice $\ell \mathbb{Z}^d$,
where $\ell$ is the lattice spacing; there is no trapping potential,
but a quantization box $[0,L]^d$ with periodic boundary conditions.
The bosonic field obeys the discrete commutation
relations
$[\hat{\psi}(\rr_1),\hat{\psi}^\dagger(\rr_2)] =
\delta_{\rr_1,\rr_2}/\ell^d.$
In the kinetic energy operator,
$\Delta$ is a discrete representation of the Laplacian on the lattice,
such that the plane wave $e^{i \kk\cdot\rr}$ is an eigenstate
of $\Delta$ with the eigenvalue $-k^2$, the discrete nature of the lattice
allowing one to restrict the values of $\kk$ to the first Brillouin
zone
\be
\mathcal{D} = [-\pi/\ell, \pi/\ell)^d.
\ee
The on-site interactions are characterized by a coupling constant 
$g_0$ adjusted to reproduce the correct value of the scattering length
$a$ in the two-body scattering problem, as detailed in \cite{Houches03}.
For the two-dimensional case one obtains
\be
g_0 = \frac{2\pi\hbar^2}{m} \frac{1}{\ln(K_2 \ell/a)}
\ \ \ \ \mbox{with} \ \ \ \ K_2 = \frac{e^{-\gamma+2G/\pi}}{\pi}.
\label{eq:g02d}
\ee
We shall find in 2D that the first correction beyond (\ref{eq:omega2})
is {\sl universal}, it depends on the interaction only through the scattering length $a$
in the zero chemical potential limit. The microscopic details
of our model, the fact that it is a lattice model
or that Catalan's constant $G=0.91596\ldots$ appears in (\ref{eq:g02d}), are 
thus not relevant in this limit.

\noindent{\sl Elimination of the condensate mode:}
We now assume that the ground state of $H$ in the thermodynamic limit
is a condensate, so that we take $d=2$ \cite{Schick,Popov} or $d=3$.
We then use Bogoliubov method to eliminate
the condensate mode and obtain a Hamiltonian for the field of
non-condensed particles. We use here a $U(1)$ symmetry preserving 
approach in the spirit of \cite{Gardiner,CastinDum}, adjusted
to the case of the grand canonical ensemble.
We split the field operator as the sum of the condensate field
and the field of the non-condensed modes:
\be
\hat{\psi}(\rr) = \phi(\rr) \hat{a}_{\mathbf{0}} + \hat{\psi}_{\rm nc}(\rr),
\label{eq:split}
\ee
where $\hat{a}_{\mathbf{0}}$ 
is the annihilation operator in the condensate mode
$\phi(\rr) = 1/L^{d/2}$.
We eliminate the condensate particle number $\hat{n}_{\kk=\mathbf{0}} = 
\hat{a}^\dagger_{\mathbf{0}}
\hat{a}_{\mathbf{0}}$ using
\be
\hat{n}_{\kk=\mathbf{0}} = \hat{N} - \hat{N}_{\rm nc},
\ee
where $\hat{N}$ is the total number of particles and
\be
\hat{N}_{\rm nc} = \sum_\rr \ell^d \hat{\psi}_{\rm nc}^\dagger \hat{\psi}_{\rm nc}
\ee
is the number of non-condensed particles.
To complete the elimination of the condensate mode, we introduce the representation \cite{epo}
\be
\hat{a}_{\mathbf{0}} = \hat{A}\, \hat{n}_{\kk=\mathbf{0}}^{1/2} \ \ \ \mbox{with} \ \ \ 
\hat{A}\equiv (1+\hat{n}_{\kk=\mathbf{0}})^{-1/2} \hat{a}_{\mathbf{0}}.
\ee
As shown in Eq.(5.40) of \cite{epo} one has the exact relations $\hat{A}\hat{A}^\dagger = 1$
and $\hat{A}^\dagger \hat{A} = 1- |\mbox{vac}_{\mathbf{0}}\rangle \langle \mbox{vac}_{\mathbf{0}}|$, where
$|\mbox{vac}_{\mathbf{0}}\rangle$ is the vacuum state for the condensate mode.
The condensate mode elimination is completed by inclusion of $\hat{A}^\dagger$ in the non-condensed
field, defining as in \cite{CastinDum} the field operator
\be
\hat{\Lambda}(\rr) = \hat{A}^\dagger \hat{\psi}_{\rm nc}(\rr),
\ee
which conserves the total particle number.
In the thermodynamic limit, the condensate mode has a
vanishing probability to be empty,
so that $\hat{A}^\dagger \hat{A} \to \hat{A}\hat{A}^\dagger = 1$, and
$\hat{\Lambda}$ and $\hat{\Lambda}^\dagger$ obey simple commutation 
relations in this limit.

In the {\sl canonical} ensemble, it remains to inject the splitting
(\ref{eq:split}) into $H$ and to eliminate the condensate mode,
finally replacing the operator $\hat{N}$ by its known value $N$.
One obtains contributions of various degrees in $\hat{\Lambda}$,
starting from degree two. The terms of degree two in $\hat{\Lambda}$ gives
the Bogoliubov Hamiltonian, the terms of higher degrees may be treated
by perturbation theory.

In the {\sl grand canonical} ensemble, however,
the chemical potential $\mu$ rather than the particle number is known;
at zero temperature, $\hat{N}$ does not fluctuate but assumes an
{\sl a priori} unknown value $N(\mu)$, a function of $\mu$
to be determined order by order in the weakly interacting limit.
To zeroth order in $\hat{\Lambda}$, the gas is a pure condensate 
and one obtains the mean-field type relation
\be
N^{(0)}(\mu) = \frac{\mu L^d}{g_0}.
\ee
It is then convenient to split $N$ as
\be
N(\mu) = N^{(0)}(\mu) + \delta N(\mu).
\ee
As we shall see, $\delta N(\mu)$ to leading order is second order
in $\hat{\Lambda}$, as the mean number of non-condensed particles
$\langle \hat{N}_{\rm nc}\rangle$.

After some calculation, neglecting unity as compared to
the condensate atom number and replacing
$\hat{N}$ with $N(\mu)$,
we obtain the desired rewriting of the Hamiltonian 
with no reference to the condensate mode:
\bea
H &\simeq & -\frac{1}{2} \mu N^{(0)}(\mu)  \nonumber \\
&+& \sum_{\rr} \ell^d \left[\hat{\Lambda}^\dagger 
\left(-\frac{\hbar^2}{2m}\Delta \right) \hat{\Lambda}
+\mu \hat{\Lambda}^\dagger \hat{\Lambda} 
+\frac{\mu}{2} \left(\hat{\Lambda}^2 + \hat{\Lambda}^{\dagger 2}\right) 
\right] \nonumber \\
&+&\frac{g_0}{L^{d/2}} \sum_\rr \ell^d  \left\{
\left[N(\mu)-\hat{N}_{\rm nc}\right]^{1/2}
\hat{\Lambda}^\dagger \hat{\Lambda}^2 +\mbox{h.c.}\right\} \nonumber \\
&+&\frac{g_0}{2 L^d} \left\{ \left[\delta N(\mu)+\hat{N}_{\rm nc}\right]^2
-4\hat{N}_{\rm nc}^2
+ \left[\delta N(\mu) -\hat{N}_{\rm nc}\right] \hat{X}\right. \nonumber \\
&+& \left. \hat{X}^\dagger \left[\delta N(\mu) -\hat{N}_{\rm nc}\right] 
\right\}
+\frac{g_0}{2} \sum_{\rr} \ell^d \hat{\Lambda}^\dagger \hat{\Lambda}^\dagger
\hat{\Lambda} \hat{\Lambda}.
\label{eq:Htrans}
\eea
We have used the fact that, in the spatially homogeneous case,
one exactly has
$\sum_\rr \ell^d \hat{\Lambda}(\rr)=0$, and we have set
\be
\hat{X} = \sum_{\rr} \ell^d \hat{\Lambda}^2.
\ee

\noindent{\sl Perturbative expansion:} 
We now expand (\ref{eq:Htrans}) in powers of $\hat{\Lambda}$.
Keeping terms up to second order in $\hat{\Lambda}$ we obtain
\begin{multline}
H_{\leq 2} = 
-\frac{1}{2} \mu N^{(0)}(\mu)
+ \sum_{\rr} \ell^d \left[\hat{\Lambda}^\dagger
\left(-\frac{\hbar^2}{2m}\Delta \right) \hat{\Lambda} \right. \\
\left. +\mu \hat{\Lambda}^\dagger \hat{\Lambda}
+\frac{\mu}{2} \left(\hat{\Lambda}^2 + \hat{\Lambda}^{\dagger 2}\right)
\right].
\end{multline}
This plays the role of the Bogoliubov Hamiltonian in the usual theory.
Its ground state energy we thus call the Bogoliubov approximation for the 
grand potential:
\be
\Omega_{\rm Bog}(\mu)=-\frac{\mu^2 L^d}{2 g_0} - \sum_{\kk \in \mathcal{D}^*} 
\epsilon_k V_k^2
\label{eq:ombog}
\ee
where we have replaced $N^{(0)}(\mu)$ by its value and we have introduced
the Bogoliubov modal amplitudes obeying
\be
U_k+V_k = \frac{1}{U_k-V_k} = \left(\frac{\hbar^2k^2/(2m)}
{2\mu+\hbar^2 k^2/(2m)}\right)^{1/4}\equiv s_k
\ee
and the corresponding Bogoliubov energies
\be
\epsilon_k = \left[\frac{\hbar^2 k^2}{2m} \left(\frac{\hbar^2 k^2}{2m}
+2\mu\right)\right]^{1/2}.
\ee
Taking minus the derivative of (\ref{eq:ombog}) with respect to $\mu$
to obtain the atom number, and using
\be
\partial_\mu (\epsilon_k V_k^2) = - V_k (U_k +V_k),
\ee
one recovers, in the thermodynamic limit, 
Eq.(152) in \cite{Mora}, and thus (\ref{eq:popov-mora}) in the limit $\ell \to 0$
\cite{precision}.

To go beyond Bogoliubov, we
collect into $H_3$ the terms of degree three in $\hat{\Lambda}$
and into $H_4$ the terms of degree four in $\hat{\Lambda}$, keeping
in mind that $\delta N(\mu)$ is to leading order of degree two,
$\delta N(\mu) = N^{(2)}(\mu) + \ldots$, so that
\bea
H_3 &=&  g_0 \left[\frac{N^{(0)}(\mu)}{L^d}\right]^{1/2} 
\sum_{\rr} \ell^d \hat{\Lambda}^\dagger (\hat{\Lambda} + \hat{\Lambda}^\dagger)
\hat{\Lambda} \\
H_4 &=&
\frac{g_0}{2 L^d} \left\{ \left[ N^{(2)}(\mu)+\hat{N}_{\rm nc} \right]^2
-4 \hat{N}_{\rm nc}^2
+ \left[N^{(2)}(\mu) -\hat{N}_{\rm nc}\right] \hat{X} \right. \nonumber \\
&+& \left. \hat{X}^\dagger \left[N^{(2)}(\mu) -\hat{N}_{\rm nc}\right] \right\}
+ \frac{g_0}{2} \sum_{\rr} \ell^d \hat{\Lambda}^\dagger \hat{\Lambda}^\dagger
\hat{\Lambda} \hat{\Lambda}.
\eea
We treat $H_4$ to first order in perturbation theory and $H_3$ to second
order, to obtain the first correction to the Bogoliubov prediction
for the grand potential:
\be
\delta \Omega = \langle H_4\rangle + \langle H_3 \frac{1}{\Omega_{\rm Bog} - 
H_{\leq 2}} H_3\rangle
\ee
where the expectation value is taken in the ground state of $H_{\leq 2}$,
that is in the vacuum of the operators $\hat{b}_\kk$ appearing in
the modal expansion
\be
\hat{\Lambda}(\rr) = L^{-d/2} \, \sum_{\kk\in \mathcal{D}^*} 
\hat{b}_\kk U_k e^{i\kk\cdot\rr} + \hat{b}_\kk^\dagger V_k e^{-i\kk\cdot\rr}.
\label{eq:modal}
\ee
To find the value of $N^{(2)}(\mu)$, we minimize $\langle H_4\rangle$ over $N^{(2)}$,
keeping in mind that $H_{\leq 2}$ and $H_3$ do not depend on $N^{(2)}$.
In the thermodynamic limit, one has $\langle \hat{N}_{\rm nc}^2\rangle
\simeq \langle \hat{N}_{\rm nc}\rangle^2$ and
$\langle \hat{N}_{\rm nc} \hat{X}\rangle \simeq 
\langle \hat{N}_{\rm nc}\rangle \langle \hat{X}\rangle$, so that
\be
N^{(2)}(\mu) \simeq - \langle \hat{N}_{\rm nc} + \frac{\hat{X}+\hat{X}^\dagger}{2}
\rangle = - \sum_{\kk\in \mathcal{D}^*} V_k (U_k + V_k).
\label{eq:N2}
\ee
The divergence of $N^{(0)}(\mu)$ when $\ell \to 0$ is removed in the combination
$N^{(0)}(\mu) + N^{(2)}(\mu)$. The resulting density is in agreement with
Eq.~\eqref{eq:popov-mora}.
One is then left with
\be
\langle H_4 \rangle \simeq - \frac{g_0}{L^d} \langle \hat{N}_{\rm nc}\rangle
\left[\langle \hat{N}_{\rm nc}\rangle + 2 \langle\hat{X}\rangle \right].
\ee
Using the modal expansion (\ref{eq:modal}) and Wick's theorem
we finally obtain after some calculation
\begin{multline}
\delta \Omega \simeq
-\frac{\mu^2}{N^{(0)}(\mu)+N^{(2)}(\mu)}
\sum_{\kk_1,\kk_3 \in \mathcal{D}^*} \left\{
\frac{V_3^2 V_1 (U_1 +s_1)}{\mu} \right. \\
+ \left. \left(1-\delta_{\kk_2,\mathbf{0}}\right)
\frac{U_1 s_2 V_3}{\epsilon_1 + \epsilon_2 + \epsilon_3}
\sum_{\sigma\in S_3} U_{\sigma(1)} s_{\sigma(2)} V_{\sigma(3)} \right\}.
\label{eq:wtis}
\end{multline}
We have introduced the vector $\kk_2 \in \mathcal{D}$ such that 
$\kk_1+\kk_2+\kk_3 \in (2\pi/l) \mathbb{Z}^d$.
The notation $U_i$, $i\in \{1,2,3\}$, stands for $U_{k_i}$.
The sum over $\sigma$ runs over the permutation group $S_3$.
For convenience we have added
$N^{(2)}(\mu)$ to $N^{(0)}(\mu)$ in the denominator of the overall factor in
(\ref{eq:wtis}), which is allowed at the present order of the calculation.

\noindent{\sl Absence of divergences in 2D:} 
We now take the thermodynamic limit, replacing sums over
$\kk$ by integrals over the domain $\mathcal{D}$ in 
(\ref{eq:wtis}). We also take the
zero lattice spacing limit $\ell \to 0$ \cite{precision} 
so that the integration domain over $\kk$
is now $\mathbb{R}^d$. 
Since $\kk_2 = -(\kk_1 + \kk_3)$ and the integrand
depends only on the moduli $k_1, k_2$ and $k_3$, see (\ref{eq:wtis}), 
we are left with a triple integral over $k_1,k_3$ 
and the angle between the vectors
$\kk_1$ and $\kk_3$. 
In 2D, we show below that this integral
converges, that is it has neither an infrared nor an ultraviolet
divergence. The first correction beyond the Bogoliubov
energy is thus {\sl universal} in 2D.
Since convergence is established,
we can resort to numerical integration.
After the change of variables $q_i = \hbar k_i/(2m\mu)^{1/2}$ and
pulling out a factor $\pi [(2m\mu)^{1/2} L/(2\pi\hbar)]^4/\mu$, we get
(\ref{eq:I}).
Summing $\Omega_{\rm Bog}$ to $\delta \Omega$ we then obtain (\ref{eq:final}).

To show the infrared convergence, we replace the integrand 
by its leading low-$k_i$ behavior: $U_i$ and $V_i$ diverge
as $1/\sqrt{k_i}$, $s_i$ vanishes as $\sqrt{k_i}$ and $\epsilon_i$ 
vanishes as $k_i$. Including the Jacobian factors $k_1$ and $k_3$
from 2D integration in polar coordinates, 
we see that the integral of the first term
in the curly brackets of (\ref{eq:wtis}) converges.
The contribution of the term due to permutation $\sigma$ scales as
\be
\frac{k_1 k_3}{k_1 + k_2 + k_3} \left(\frac{k_2}{k_1 k_3}\right)^{1/2}
\left(\frac{k_{\sigma(2)}}{k_{\sigma(1)} k_{\sigma(3)}}\right)^{1/2}
< 1,
\ee
so its integral over $k_1$ and $k_3$ is also convergent.

For the ultraviolet convergence, 
the full reasoning is rather long \cite{aide}, so we 
give a simplified explanation. We approximate each term in the integrand 
in (\ref{eq:wtis}) by its leading high-$k_i$ behavior, $U_i$ and $s_i$
tending to unity, $V_i$ vanishing as $1/k_i^2$ and $\epsilon_i$ diverging 
as $k_i^2$. In the sum over $\sigma$, the terms with $\sigma(3)\neq 3$
are not dangerous. E.g.\ for $\sigma(3)=1$,
a factor $V_1 V_3$ appears, and including the Jacobian factors,
one obtains a contribution scaling as
\be
\frac{k_1 k_3}{k_1^2 + k_2^2 + k_3^2}\times  \frac{1}{k_1^2 k_3^2} <
\frac{1}{2 k_1^2 k_3^2},
\ee
so that the resulting
double integral over $k_1$ and $k_3$ is convergent at infinity.
The dangerous terms in the sum over $\sigma$ 
thus correspond to $\sigma(3)=3$: 
the factor $V_3^2$ ensures convergence of the integral over $\kk_3$
over a  $\kk_1$-independent range~$\sim (m\mu/\hbar^2)^{1/2}$.
At large $\kk_1$, the energy denominator $\epsilon_1+\epsilon_2+\epsilon_3$
approaches $2 \epsilon_1$.
Then, from the asymptotic relation $V_1\simeq -\mu/(2\epsilon_1)$, 
we see that the two dangerous contributions coming from the permutations
with $\sigma(3)=3$ exactly compensate with the first term
$\simeq 2 V_3^2  V_1/\mu$ in the curly brackets of (\ref{eq:wtis}),
which avoids an ultraviolet divergence of $\delta \Omega$.

In conclusion, we have calculated analytically and in a systematic
way the first correction
to the Bogoliubov prediction for the ground state grand potential
of a 2D weakly interacting Bose gas. 
We find that this correction is {\sl universal}, depending on the interaction potential
through the scattering length only.
It allows to describe analytically the not extremely weakly
interacting regime, and contrarily to other analytical works,
we obtain a prediction for the ground state energy
in excellent agreement with the numerical results
of \cite{Giorgini,Astra} over the range where the results of 
\cite{Giorgini,Astra} are model independent.

This work was stimulated by discussions with Elliot Lieb and Jakob Yngvason.
The group of Y.C. is a member of IFRAF.



\begin{thebibliography}{99}

\bibitem{Jean_BKT}
Z. Hadzibabic {\sl et al.},
Nature {\bf 441}, 1118 (2006).

\bibitem{Phillips_BKT}
P. Clad\'e {\sl et al.}, 
arXiv:0805.3519 (2008).

\bibitem{revue1}
A. Posazhennikova, Rev. Mod. Phys. {\bf 78}, 1111 (2006).

\bibitem{revue2}
I. Bloch, J. Dalibard, and W. Zwerger, 
Rev. Mod. Phys. {\bf 80}, 885 (2008).

\bibitem{Schick}
M. Schick, Phys. Rev. A {\bf 3}, 1067 (1971).

\bibitem{Popov} 
V.N. Popov, Theor. Math. Phys. {\bf 11}, 565 (1972).

\bibitem{Lieb}
E. Lieb and J. Yngvason, J. Stat. Phys. {\bf 103}, 509 (2001).

\bibitem{Giorgini} 
S. Pilati, J. Boronat, J. Casulleras and S. Giorgini, 
Phys. Rev. A {\bf 71}, 023605 (2005).

\bibitem{Astra}
G.E. Astrakharchik {\sl et al.}, 
arXiv:0812.3844v1 (2008).

\bibitem{Mora}
C. Mora and Y. Castin, Phys. Rev. A {\bf 67}, 053615 (2003).

\bibitem{Cherny} 
A.Yu. Cherny and A.A. Shanenko, Phys. Rev. E {\bf 64}, 027105 (2001).

\bibitem{Andersen}
J.O. Andersen, Eur. Phys. J. B {\bf 28}, 389 (2002).

\bibitem{Pricoupenko}
L. Pricoupenko, Phys. Rev. A {\bf 70}, 013601 (2004).

\bibitem{info_Popov}
We found that the quantity $\epsilon_0$ appearing in Popov's theory
is simply  $\epsilon_0 = 
4\hbar^2/(m a^2 e^{2\gamma})$. 
This is obtained
by comparing the low-$k$ expression of the two-body $T$ matrix
in Eq.(3.2) of \cite{Popov} with Eq.(160) of \cite{Mora}.

\bibitem{Wu}
T. T. Wu, Phys. Rev. {\bf 115}, 1390 (1959).

\bibitem{param}
We used the relations $m\mu a^2/\hbar^2 = 
t^2[1-1/\ln(q t)]$ and $2\pi \rho a^2 = t^2[-\ln(qt)+1/2]$
with $q=\exp(\gamma)/2$ and $0<t<1/q$ (L. Pricoupenko, private
communication).

\bibitem{Houches03}
Y. Castin, 
J. Phys. IV (France) {\bf 116}, 89 (2004).

\bibitem{Gardiner}
C.W. Gardiner, Phys. Rev. A {\bf 56}, 1414 (1997).

\bibitem{CastinDum}
Y. Castin, R. Dum, Phys. Rev. A {\bf 57}, 3008 (1998).

\bibitem{epo}
P. Carruthers, M. Nieto, Rev. Mod. Phys. {\bf 40}, 411 (1968).

\bibitem{aide}
Setting $q_i= \hbar  k_i /(2m\mu)^{1/2}$,
we used e.g.\ $| 2 s_2 V_2| = |s_2^2-1| \leq 2/(2+q_2^2)$
and $(\epsilon_1-\epsilon_2)/(q_1^2-q_2^2) \leq 1+2/(q_1^2+q_2^2)$.

\bibitem{precision}
More precisely, in 2D, our expansion in powers of $\hat{\Lambda}$ 
is an expansion in powers of $\epsilon(\mu)$ for a fixed
value of $\eta(\mu)=\ln(\xi/\ell)\gg 1$, where $\hbar^2/(m\xi^2)=\mu$. 
The Bogoliubov method indeed relies on a Born expansion \cite{LiebLiniger} (here in powers
of $g_0$) of the scattering
amplitude $f_k$ for $k\simeq 1/\xi$. From Eq.(167)
of \cite{Mora} the small parameter for the Born
expansion is $2\eta\epsilon$.
This small parameter is explicitly obtained in our approach, from
the requirement $N^{(2)}\ll N^{(0)}$:
Estimating from (\ref{eq:N2}) $N^{(2)}\simeq \eta L^2 m \mu/(2\pi\hbar^2)$ 
for $\ell\ll \xi$, we get $N^{(2)}/N^{(0)}\simeq 2\eta \epsilon/(1+2\eta \epsilon)$.
One then takes the limit $\eta\to +\infty$ 
in each coefficient of the expansion of $\Omega$ in powers of $\epsilon$.
This limit is exponentially fast approached in $\eta$:
we find that $\eta=7$ is more than large enough,
it gives the value of $I$ in (\ref{eq:I}) at the $10^{-5}$ level.

\bibitem{LiebLiniger}
E.H. Lieb and W. Liniger, Phys. Rev. {\bf 130}, 1605 (1963).

\end{thebibliography}
\end{document}